\documentclass[preprint,12pt]{elsarticle}
\usepackage{tablefootnote}
\usepackage[graphicx]{realboxes}
\usepackage{varwidth}

\usepackage{amssymb}
\usepackage{lineno}
\biboptions{sort&compress}

\usepackage{appendix}
\usepackage{xcolor}

\newcommand\dele[1]{\textcolor{blue}{ #1}}
\renewcommand\dele[1]{ }

\newcommand\JCAP{JCAP }
\newcommand\Sc{Science }
\newcommand\Ap{Astropart.~Phys. }
\newcommand\ApJ{Astrophys.~J. }
\newcommand\AaA{Astron.~Astrophys. }
\newcommand\EPJ{Eur.~Phys.~J. }
\newcommand\nima{Nucl.~Instrum.~Methods~A }
\newcommand\nim{Nucl.~Instrum.~Methods }
\newcommand\pr{Phys.~Rev. }
\newcommand\mnras{MNRAS }

\newcommand\beq{\begin{equation}}
\newcommand\beql[1]{\begin{equation} \label{#1}}
\newcommand\eeq{\end{equation}}
\newcommand\ben{\begin{eqnarray}}
\newcommand\een{\end{eqnarray}}
\newcommand\bea{\begin{array}}
\newcommand\eea{\end{array}}
\newcommand\bem{\begin{displaymath}}
\newcommand\eem{\end{displaymath}}

\newcommand\eqa[1]{Eq.(\ref{#1})}
\newcommand\eqb[1]{Eqs.(\ref{#1})}
\newcommand\eqc[1]{(\ref{#1})}

\newcommand\fig[1]{Fig.\ref{#1}}
\newcommand\figs[1]{Figs.\ref{#1}}
\newcommand\figg[1]{\ref{#1}}
\newcommand\tab[1]{Table~\ref{#1}}

\newcommand\sct[1]{Section~\ref{#1}}
\newcommand\sctw[2]{Sections~\ref{#1}~and~\ref{#2}}

\newcommand\app[1]{Appendix~\ref{#1}}
\renewcommand\app[1]{\ref{#1}}

\newcommand\qqa{\quad}
\newcommand\qqb{\qquad}
\newcommand\qqc{\qquad \qquad}

\newcommand\wsa{\vspace{-0.5cm}}

\newcommand\wse{\vspace{-3.5cm}}

\newcommand\Ssum[2]{\sum \limits_{#1}^{#2}}

\newcommand\Iint[2]{\int \limits_{#1}^{#2}}
\newcommand\dif[1]{{\rm \,d}{#1}}

\newcommand\dg{\mbox{$^{\circ}$}}

\newcommand\EeV{ \, {\rm EeV} }
\newcommand\AMpc{ \, {\rm Mpc} }
\newcommand\kmSsry{ \, \mbox{\rm km$^{2}$\,sr\,y } }

\newcommand\non{n_{\rm on} }
\newcommand\noff{n_{\rm off} }
\newcommand\mon{\mu_{\rm on} }
\newcommand\moff{\mu_{\rm off} }
\newcommand\ms{\mu_{\rm s} }
\newcommand\mb{\mu_{\rm b} }

\newcommand\Non{N_{\rm on} }
\newcommand\Noff{N_{\rm off} }

\newcommand\Bii{{\rm Bi} }
\newcommand\NBii{{\rm NBi} }

\newcommand\Poo{{\rm Po} }
\newcommand\Gaa{{\rm Ga} }
\newcommand\Be{{\rm B} }
\newcommand\Bg{{\rm B}_{\rm{g2}} }

\newcommand\SLM{S_{\rm LM} }
\newcommand\SBa{S_{\rm B}}

\newcommand\CL{{\rm CL} }

\newcommand\ommr{\omega_{+} }
\newcommand\omml{\omega_{-} }

\newcommand\fbet{f_{\beta} }
\newcommand\fbetp{f^{+}_{\beta} }

\newcommand\fdel{f_{\delta} }

\newcommand\fom{f_{\omega} }
\newcommand\fomp{f^{+}_{\omega} }

\newcommand\fjj{f_{j} }
\newcommand\fjjp{f^{+}_{j} }

\newcommand\fdelb{h_{\delta} }

\newcommand\fbetb{h_{\beta} }

\newcommand\fomb{h_{\omega} }

\newcommand\ftau{f_{\tau}}
\newcommand\ftaup{f_{\tau}^{+}}
\newcommand\ftaub{h_{\tau} }

\newcommand\ftaubp{h^{+}_{\tau} }

\newcommand\atau{\lambda_{\tau}}

\newcommand\Pm{P^{-} }
\newcommand\Pp{P^{+} }

\newcommand\Pmb{R^{-} }
\newcommand\Ppb{R^{+} }

\newcommand\aon{\frac{\alpha}{1+\alpha} }

\newcommand\ssp{s_{p} }
\newcommand\ssq{s_{q} }
\newcommand\gp{\gamma_{p} }
\newcommand\gq{\gamma_{q} }
\newcommand\gqa{\frac{\gamma_{q}}{\alpha} }
\newcommand\gqb{\left( \frac{\gamma_{q}}{\alpha} \right) }
\newcommand\gqc{\frac{\gp}{\gq}}

\newcommand\gpqb{\gamma_{p} + \frac{\gamma_{q}}{\alpha} }
\newcommand\gpqc{\frac{\alpha \gamma_{p}}{\gamma_{q}} }

\newcommand\rrha{\frac{1}{1+\rho} }
\newcommand\rrhb{\frac{\rho}{1+\rho} }

\renewcommand\gqc{\gp / \gq}
\renewcommand\gpqc{\alpha \gamma_{p} / \gamma_{q} }
\newcommand\ja{j_{1}}
\newcommand\jb{j_{2}}
\newcommand\deltaa{\delta_{1}}
\newcommand\deltab{\delta_{2}}
\newcommand\betaa{\beta_{1}}
\newcommand\betab{\beta_{2}}

\newcommand\fdelaa{f_{\delta_{1}}}
\newcommand\fdelbb{f_{\delta_{2}}}
\newcommand\fbetaa{f_{\beta_{1}}}
\newcommand\fbetbb{f_{\beta_{2}}}

\newcommand\nona{n_{{\rm on}_{1}}}
\newcommand\nofa{n_{{\rm off}_{1}}}

\newcommand\mba{\mu_{{\rm b}_{1}}}
\newcommand\eona{a_{1}}

\newcommand\xxaa{x_{1}}
\newcommand\alphaa{\alpha_{1}}
\newcommand\gpaa{\gamma_{p_{1}}}
\newcommand\sspa{s_{p_{1}}}

\newcommand\nonb{n_{{\rm on}_{2}}}
\newcommand\nofb{n_{{\rm off}_{2}}}

\newcommand\mbb{\mu_{{\rm b}_{2}}}
\newcommand\eonb{a_{2}}

\newcommand\xxbb{x_{2}}
\newcommand\alphab{\alpha_{2}}
\newcommand\gpbb{\gamma_{p_{2}}}
\newcommand\sspb{s_{p_{2}}}

\newcommand\AvB{\mbox{$A$-$B$} }
\newcommand\AvC{\mbox{$A$-$C$} }
\newcommand\AvD{\mbox{$A$-$D$} }
\newcommand\BvC{\mbox{$B$-$C$} }

\newcommand\CvD{\mbox{$C$-$D$} }
\newcommand\EvF{\mbox{$E$-$F$} }

\newcommand\YY{Y}
\newcommand\Yon{Y_{\rm on}}

\newcommand\taua{\tau_{1}}
\newcommand\taub{\tau_{2}}
\newcommand\DDa{D_{1}}
\newcommand\DDb{D_{2}}

\newcommand\NN{N}

\newcommand\midd{\! \mid \!}

\journal{Nuclear Instruments and Methods A}

\begin{document}

%------------------------------------------------------------------
\begin{frontmatter}

\title{A Bayesian on-off analysis of cosmic ray data}


\author[nos01]{Dalibor Nosek}
\ead{nosek@ipnp.troja.mff.cuni.cz}
\author[nos02]{Jana Noskov\'a}
\address[nos01]{Charles University, Faculty of Mathematics and Physics,
Prague, Czech Republic}
\address[nos02]{Czech Technical University, Faculty of Civil Engineering,
Prague, Czech Republic}

%------------------------------------------------------------------
\begin{abstract}
We deal with the analysis of on-off measurements designed for 
the confirmation of a weak source of events whose presence 
is hypothesized, based on former observations.
The problem of a small number of source events that are masked by 
an imprecisely known background is addressed from a Bayesian point 
of view.
We examine three closely related variables, the posterior 
distributions of which carry relevant information about various 
aspects of the investigated phenomena. 
This information is utilized for predictions of further observations, 
given actual data.
Backed by details of detection, we propose how to quantify 
disparities between different measurements.
The usefulness of the Bayesian inference is demonstrated on examples 
taken from cosmic ray physics. 
\end{abstract}
%------------------------------------------------------------------

%------------------------------------------------------------------
\begin{keyword}
Bayesian inference \sep
On-off problem \sep
Source detection \sep
Cosmic rays
\end{keyword}

\end{frontmatter}

%------------------------------------------------------------------
%\begin{linenumbers}
%\linenumbers

%------------------------------------------------------------------
\section{Introduction}
\label{Sec01}

The search for new phenomena often yields data that consists of 
a set of discrete events distributed in time, space, energy or 
some other observables.
In most cases, source events associated with a new effect are hidden 
by background events, while these two classes of events cannot be 
distinguished in principle.
Such a search can be accomplished with an on-off measurement by
checking whether the same process of a constant but unknown intensity 
may be responsible for observed counts in the on-source region, where 
a new phenomenon is searched for, and in the reference off-source region,
where only background events contribute. 
Any inconsistency between the numbers of events collected in these zones, 
when they are properly normalized, then indicates the predominance of 
a source producing more events in one explored region over 
the other.

In this study, we focus on the problems which are often encountered 
when searching for cosmic ray sources while detecting rare events.
Characteristics of possible sources are usually proposed 
based on analysis of a test set of observed data.
Then, further observations are to be conducted in order 
to examine the presence of a source or to improve conditions 
for its verifications.  
But, due to unknown phenomena, the outcome is always uncertain
which calls, first, for as less as possible initial assumptions 
about underlying processes and, second, for the quantification
of disparities between observations with the option to correct 
for experimental imperfections. 

In order to satisfy the first condition, we follow our previous 
analysis of on-off measurements formulated within the Bayesian
setting~\cite{Nos01}.
Unlike other Bayesian 
approaches~\cite{Kno01,Cas01,Hel01,Hel02,Pro01,Pro02,Gil01,Gre01}, 
we handle the source and background processes on an equal footing.
This option provides us with solutions that are minimally affected 
by external presumptions. 
In order to track the behavior of a signal registered in a selected 
on-source region, we utilize variables with the capability to assess 
the consistency between on-off measurements.
Specifically, giving the net effect, the difference variable~\cite{Nos01} 
is well suited for estimating source fluxes if exposures are known.
In case of stable or at least predictable background rates, 
we eliminate the effect of exposures by using fractional 
variables which reveal relatively the manifestation of a source. 
For example, the time evolution of a given source, if still observed 
in the same way, is easily examined by the ratio of the on-source rate 
to the total rate. 
In a more general case, we employ the on-source rate expressed 
in terms of the rate deduced from the background. 
In summary, we receive posterior distributions of different
variables that include what is available from measurements, while 
providing us with all kinds of estimates, as traditionally communicated, 
and allowing us to make various observation-based predictions.

Related to the on-off issue, the Bayesian inference provides solutions 
in the case of small numbers, including the null experiment or 
the experiment with no background, when classical methods based on 
the asymptotic properties of the likelihood ratio 
statistic~\cite{Wil01,Lim01,Cou01,Cow01} are not easily applicable.
Also, there are no difficulties with the regularity conditions 
of Wilks' theorem, with unphysical likelihood estimates
or with the discreteness of counting experiments, in general,  
see e.g. Refs.~\cite{Fel01,Cou02,Rol01,Alg01}.
On the other hand, the subjective nature of Bayesian reasoning, 
often mentioned as its disadvantage, may be at least partially 
eliminated by using a family of uninformative prior options.

The proposed method is suitable for experiments searching 
for rare events in which the observational conditions may not be adjusted 
optimally, with little opportunity for repeating measurements conducted 
under exactly the same conditions.
Besides searches for possible sources of the highest energy 
cosmic rays, see e.g. Refs.~\cite{Aug01,Aug02,Aug03,Aug04,Aug05},
examples include observations of peculiar sources which
exhibit surprising temporal or spectral behavior.
Another class of observations comprises searches for events
accompanying radiation from transient sources that have been
identified in different energy ranges. 
The identification of the properties of very-high-energy $\gamma$-rays
associated with observed gamma-ray bursts belongs to this class
of problems~\cite{Nos01,Kno01,Cas01}.

The structure of this paper is as follows.
Our formulation of the Bayesian approach to the on-off 
problem is described in~\sct{Sec02}, complemented by five 
Appendices.
Further details about our approach can be found in Ref.~\cite{Nos01}.
In~\sct{Sec02a} we summarize how to store experimental information 
by using appropriate on-off variables. 
Two ways to examine possible inconsistencies in independent 
observations are proposed in~\sctw{Sec02b}{Sec02c}.
Several realistic examples taken from cosmic ray physics 
are presented and discussed in~\sct{Sec03}.
The paper is concluded in~\sct{Sec04}.

%------------------------------------------------------------------
\section{Bayesian inferences from on-off experiment}
\label{Sec02}

In the on-off experiment, two kinds of measurements are 
collected in order to validate a source signal immersed 
in background.
The number of on-source events, $\non$, is recorded in a signal 
on-source region, while the number of off-source events, $\noff$, 
detected in a background off-source zone serves as a reference 
measurement.
The on- and off-source counts are modeled as discrete random variables 
generated in two independent Poisson processes with unknown on- and 
off-source means, $\mon$ and $\moff$, 
i.e. $\non \sim \Poo(\mon)$ and $\noff \sim \Poo(\moff)$.
The relationship between the on- and off-source zone is ensured 
by the ratio of on- and off-source exposures $\alpha > 0$.

In the Bayesian approach, for on- and off-source means we adopted 
a family of prior distributions conjugate to the Poisson sampling 
process~\cite{Nos01}.  
This family consists of Gamma distributions, 
i.e. 
\beql{Q01}
\mon \sim \Gaa(\ssp,\gp-1), \qqc
\moff \sim \Gaa(\ssq,\gq-1),
\eeq
where $\ssp >0$ and $\ssq > 0$ are prior 
shape parameters, and the prior rate parameters $\gp > 1$ and $\gq > 1$.
It includes several frequently discussed options, i.e. scale 
invariant, uniform, as well as Jeffreys' prior distributions.
After the on-off measurement has been conducted, when $\non$ and 
$\noff$ counts were registered independently in the on- and 
off-source regions, using~\eqa{Q01} we obtain independent posterior 
distributions
\beql{Q02}
(\mon \midd \non) \sim \Gaa(p,\gp), \qqc 
(\mb \midd \noff) \sim \Gaa(q,\gqa),
\eeq
where $\mb = \alpha \moff$ denotes the expected background rate in 
the on-source zone and $p = \non+\ssp$ and $q=\noff+\ssq$. 
For more details see Ref.~\cite{Nos01}.

We recall that our next steps diverge from the traditional 
treatment.
In order to assess what is observed, we define suitable on-off 
variables by combining the on- and off-source means, assuming 
that the underlying processes are independent.
From the Bayesian perspective, this choice is motivated by the fact 
that, according to Jeffreys' rule, the joint prior distribution is 
separable in the on- and off-source means~\cite{Nos01,Kno01}. 
Furthermore, as in classical statistical 
approaches~\cite{Lim01,Wil01,Cou01,Cow01,Fel01,Cou02,Rol01}, 
the proposed option allows us to obtain adequate results regardless 
of in which of the two zones the source effects are 
revealed~\cite{Nos01,Pro02}.

%------------------------------------------------------------------
\subsection{On-off variables}
\label{Sec02a}

In our previous work~\cite{Nos01}, we focused on the properties 
of the difference between the on-source and background means, 
$\delta = \mon - \mb$, using maximally uninformative joint 
distributions, as dictated by the principle of maximum entropy.
In this section, we briefly recapitulate our previous result and
introduce other on-off variables that equally well 
describe the on-off problem. 

Under the transformation $\delta = \mon - \mb$, with a real valued 
domain, while keeping $\mb = \alpha \moff$ unchanged and  
marginalizing over $\mb$, the probability density function of 
the difference is (for details of our notation see Ref.~\cite{Nos01})
\beql{Q03}
\fdel(x) = 
\frac{\gp^{p} \gqb^{q}}{\Gamma(p)} \
e^{-\gp x} x^{p+q-1} \
U(q,p+q, \eta x), \qqa
x \ge 0,
\eeq
\beql{Q04}
\fdel(x) = 
\frac{\gp^{p} \gqb^{q}}{\Gamma(q)} \
e^{\gqa x} (-x)^{p+q-1} \
U(p,p+q,-\eta x), \qqa
x < 0,
\eeq
where $p = \non+\ssp$, $q=\noff+\ssq$, $\eta = \gpqb$, 
$\Gamma(a)$ stands for the Gamma function 
and $U(a,b,z)$ is the Tricomi confluent hypergeometric 
function~\cite{Olv01}. 
Exhaustive discussion concerning this distribution can be found in 
Ref.~\cite{Nos01}, where also some special cases ($\gp = \gq \to 1$) 
based on uninformative prior distributions, scale invariant 
($\ssp = \ssq \to 0$), Jeffreys' ($\ssp = \ssq = \frac{1}{2}$) 
and uniform ($\ssp = \ssq = 1$) options, are described. 

The difference $\delta$ yields information about the source flux. 
The posterior distribution of the source flux is obtained by 
a scale transformation, i.e. $j = \delta / a$ where $a = \aon A$ 
is the exposure of the on-source zone and $A$ denotes 
the integrated exposure of the on-off experiment, both considered 
as constants.

A similar picture is obtained with the ratio of the on-source and 
background means ($\mb = \alpha \moff$) 
\beql{Q05}
\beta = \frac{\mon}{\mb}, \qqc 
\beta \ge 0.
\eeq
This variable represents the intensity registered in the on-source 
region expressed in terms of the background intensity,
i.e. $\beta \le 1$ when no source is present in the on-source zone.
The ratio $\beta$ obeys the generalized Beta distribution of the second 
kind~\cite{Mcd01},
$\beta \sim \Bg(p,q,\rho)$ where $p = \non+\ssp$, $q=\noff+\ssq$ and
$\rho = \gpqc$, with the probability density function 
\beql{Q06}
\fbet(x) = 
\frac{\rho^{p}}{B(p,q)} \frac{x^{p-1}}{(1+\rho x)^{p+q}}, \qqc
x \ge 0,
\eeq
where $B(a,b)$ is the Beta function~\cite{Olv01}.
This posterior distribution was obtained after the transformation 
$\beta = \mon / \mb$ while treating $\mon$ and $\mb$ as independent 
variables (see~\eqa{Q02}) and keeping $\mb$ unchanged, with 
the Jacobian $J = \mb$, and marginalizing over $\mb$.

In a special case, using the uniform prior distributions for the on- 
and off-source means, i.e. $\gp = \gq \to 1$ and $\ssp = \ssq = 1$, 
and assuming that the on-off data were registered in the regions of 
the same exposure, when $\rho = \alpha = 1$, the posterior distribution 
for the ratio $\beta$ written in~\eqa{Q06} reduces to the result given 
originally in Ref.~\cite{Hel02}. 
Assuming $\gp = \gq \to 1$ and $\alpha = 1$, i.e. $\rho = 1$, the 
result presented in Eq.(13) in Ref.~\cite{Pro01} is obtained.

In some cases, it may be appropriate to use a variable 
\beql{Q07}
\omega = \frac{\mon}{\mon + \moff}, \qqc 
\omega \in \langle 0, 1 \rangle,
\eeq 
that represents the fraction 
of the total intensity registered in the on-source zone.
Considering that $\omega = \alpha \beta / (1 + \alpha \beta)$,
we recover from~\eqa{Q06} that the probability density function 
of the proportion $\omega$ is  
\beql{Q08}
\fom(x) = 
\frac{\kappa^{p}}{B(p,q)} \, 
\frac{x^{p-1} (1-x)^{q-1}}{\left[ 1 + (\kappa-1) x \right]^{p+q}}, \qqc
x \in \langle 0, 1 \rangle,
\eeq
where $p = \non+\ssp$, $q=\noff+\ssq$ and $\kappa = \gqc$ 
is the ratio of the prior rate parameters. 
In this case, equally intensive on- and off-source processes 
($\mon = \mb$) are described by a balance value of $\omega = \aon$.

Note that any Bayesian statement based on the probabilities inferred 
from the above derived distributions is independent of the prior rate 
parameters when $\gp = \gq$ and thus $\rho = \alpha$.
For $\gp = \gq$, we even have that the proportion $\omega$ obeys 
the Beta distribution, i.e. $\omega \sim \Be(p,q)$.
This widely used option also follows from using the prior Beta
distributions conjugate to the binomial sampling process, 
i.e. prior $\omega \sim \Be(\ssp,\ssq)$.
In the context of on-off measurements, the classical analysis of 
the binomial proportion is discussed in Refs.~\cite{Cou01,Cou02}, 
for example.
Point estimates of the proportion $\omega$ are traditionally 
used in the analysis of directional data in cosmic ray physics, 
see e.g. Ref.~\cite{Aug01,Aug02,Aug03,Aug05,Tar01,Tar02}.

The proposed Bayesian solutions to the on-off problem have 
other interesting features.
Unlike traditional 
approaches~\cite{Kno01,Cas01,Hel01,Hel02,Pro01,Pro02,Gil01,Gre01}, 
we treat the on- and off-source processes as independent.
Hence, our posterior distributions are maximally noncommittal about 
missing information on the relationship between these processes.
Moreover, receiving information separately from the on- and off-source 
observations, the on-off problem is examined without a predetermined 
assumption in which zone the source is to be searched for~\cite{Nos01}. 
Thus, any detected imbalance will lead to the same conclusion 
notwithstanding the region where more activity is expected~\cite{Nos01}.
Note that most classical test statistics relevant to the on-off problem 
possess the same property~\cite{Cou02,Rol01,Lim01}.

Other technical details are summarized in Appendices. 
In~\app{App01} we show that all three on-off variables provide 
the same probability of the source absence in the on-source zone. 
Note, however, that the fractional variables $\beta$ or $\omega$,
which are easier to handle, do not substitute for 
the difference $\delta$. 

A way how to determine the shortest credible intervals for 
the on-off variables is described in~\app{App02}. 
In~\app{App03} we show how to modify Bayesian solutions, when 
a source is known to be present in the on-source zone. 
Similar solutions are also obtained in often adopted schemes, 
whereby source and background parameters are treated as 
independent variables~\cite{Kno01,Cas01,Pro01,Pro02,Gil01,Gre01}.
In~\app{App04} we present Bayesian solutions for cases when 
background rates are known with sufficient precision. 

%------------------------------------------------------------------
\subsection{Waiting for next events}
\label{Sec02b}

Current experiments collecting rare events raise interest for 
predictions based on previous observations.
Typically, we want to know how many events must be registered 
in a subsequent experiment in order to identify a given number 
of events in a selected on-source zone, while relying on previous 
data collected under the same conditions with the same instrument. 
This issue is solved by constructing a relevant predictive 
distribution.

According to previous considerations, we assume that the numbers 
of on- and off-source events registered in a new experiment up to 
and including time $t$ are generated in two independent Poisson 
processes $\{ \Non(t); t \ge 0 \}$ and $\{ \Noff(t); t \ge 0 \}$
with respective rates $\mon$ and $\moff$, i.e. among others, 
$\Non(t) \sim \Poo(\mon t)$ and $\Noff(t) \sim \Poo(\moff t)$.  
Hence, we know that events of the merged Poisson process 
$\{ \NN(t) = \Non(t) + \Noff(t); t \ge 0 \}$, 
$\NN(t) \sim \Poo(\mu t)$ where $\mu = \mon + \moff$, 
arrive into the on-source zone with the probability 
$\omega = \mon / \mu$ independently of each other 
and independently of their arrival times, see e.g. Ref.~\cite{Tij01}.
Consequently, if the total number of events $n > 0$ is
collected up to time $t$, the corresponding number of on-source events, 
$\Yon = (\Non(t) \midd \NN(t) = n)$,
has a binomial distribution with parameters $n$ and $\omega$, i.e. 
$\Yon \sim \Bii(n,\omega)$.
We also know that the total number of events recorded until 
a predefined number $k > 0$ of events arrive into the on-source zone, 
$\YY = (\NN(t) \midd \Non(t) = k, 
\mbox{the on-source event is the last one})$,  
has a shifted negative binomial distribution (waiting time
distribution) with parameters $k$ and $\omega$, i.e. 
$\YY \sim \NBii(k,\omega)$ with support $n=k,k+1,\dots$, 
see e.g. Ref.~\cite{Jon01}.

Further, we ask for the probability $p_{n,k}(\omega)$ that more than 
$n$ events in total are collected before the $k$-th on-source event 
is registered if, as justified above, events are switched independently 
between on- and off-source zones with the probability $\omega$. 
We obtain ($k >0$ and $n=k,k+1,\dots$)
\beql{Q09}
p_{n,k}(\omega) = 
P( \YY > n \midd \omega ) = 
P( \Yon < k \midd \omega ) = 
\Ssum{i=0}{k-1} {n \choose i} 
\omega^{i} ( 1 - \omega )^{n-i},
\eeq
where we use the relation between the negative binomial variable 
$\YY$ and the binomial variable $\Yon$, see e.g.~Eq.(5.31) in 
Ref.~\cite{Jon01}.
This way, \eqa{Q09} gives the probability of the waiting time 
for the $k$-th on-source event when the time is measured 
in terms of the total number of collected events $n$.

In order to determine the chances of identifying on-source 
events in a new series of observations, we need to be informed 
about the binomial parameter $\omega$.
We use the fact that, in the Bayesian concept, the information 
on future measurements is contained in the posterior 
predictive distribution of unobserved observations, conditional 
on the already observed data.
This distribution is obtained by marginalizing the distribution 
of the new data, given parameters, over the posterior distribution 
of parameters, given the previous data, accounting thus for uncertainty 
about involved parameters.

Since the Poisson processes guarantee that the new and 
old observations in disjoint time intervals are independent, 
when conditioned on parameters $\mon$ and $\moff$, or, 
equivalently, on $\mu = \mon + \moff$ and $\omega = \mon / \mu$, 
and since the waiting time probability given in~\eqa{Q09} 
is independent of $\mu$, we can write
\beql{Q10}
P( \YY > n, \mu, \omega \midd D) = 
P( \YY > n \midd \omega) p(\mu, \omega \midd D),
\eeq
where $D = (\non, \noff)$ denotes the old on-off data and 
$p(\mu, \omega \midd D)$ is the joint posterior distribution 
of $\mu$ and $\omega$ which is obtained via Bayes' rule using 
the prior distributions for $\mon$ and $\moff$ in~\eqa{Q01}.
Hence, by marginalizing over $\mu$ and $\omega$, we obtain 
from~\eqb{Q09} and~\eqc{Q10} that, in the new data set, 
the waiting time for the $k$-th on-source event exceeds $n$ 
with the probability 
\beql{Q11}
P_{n,k} =
\Iint{0}{1} 
\left[ \Iint{0}{\infty} 
P( \YY > n, \mu = y, \omega = x \midd D) \dif{y} 
\right] \dif{x} = 
\Iint{0}{1} p_{n,k}(x) \fom(x) \dif{x},
\eeq
where 
$\fom(x) = p(\omega = x \midd D) = 
\Iint{0}{\infty} p(\mu = y, \omega = x \midd D) \dif{y}$ 
is the posterior distribution of the proportion $\omega$ given 
in~\eqa{Q08}. 
In particular, assuming that $\omega \sim B(p,q)$ for $\gp = \gq$ 
($\kappa = 1$) where $p = \non + \ssp$ and $q = \noff + \ssq$
are known from the previous measurement, it follows that 
\beql{Q12}
P_{n,k} = 
\Iint{0}{1} p_{n,k}(x) \fom(x) \dif{x} =
\Ssum{i=0}{k-1} {n \choose i}
\frac{B(p+i,q+n-i)}{B(p,q)}.
\eeq
Here, the Beta functions are replaced by the incomplete Beta 
functions, $B(a,b) \to B_{\frac{1}{1+\alpha}}(a,b)$, 
if a source is considered to be present in the on-source zone, 
see~\app{App03}.  

The application of this result to the new data allows us 
to assess the consistency between subsequent observations. 
Consider that $n$ new events in total are registered until 
the $k$-th new event arrives into the on-source zone, while 
the previous data has been processed.
We know that the probability of the new observation is $P_{n,k}$ 
provided the new and old data are generated in the counting 
model described above.
In the classical sense, it means that our initial assumptions 
are not valid at a level of confidence $\CL < 1 - P_{n,k}$.
Hence, at this level of confidence, our data-driven model fails 
to describe what has been measured and we conclude that, 
besides other possibilities, the new data may indicate 
a smaller on-source signal or a larger background rate 
than would correspond to the previous measurement.

%------------------------------------------------------------------
\subsection{Comparison of on-off measurements}
\label{Sec02c}

In this Section we address the question of how to compare 
two independent on-off measurements.
Our goal is to quantify statistically which of the measurements 
indicate a more intense emitter, while relying on information 
about observations contained in the posterior distributions 
of on-off variables.
Besides sequential measurements performed under similar 
conditions, we also admit experiments conducted with different 
equipments, for example, when different sources in different 
spatial, time or energy ranges are observed.

We assume that two independent on-off observations, marked by indices 
$1$ and $2$, were collected and processed by the method described 
in~\sct{Sec02a}.
Depending on what we want to examine, we choose one type of 
the on-off variable.
The relationship between the two Bayesian outputs 
is quantified by the probability 
$P( \taua < A \, \taub \midd \DDa, \DDb )$ where $\taua (\taub)$ 
is a suitable on-off variable ($\tau = \delta, \beta$ or $\omega$) 
for the first (second) measurement 
and $\DDa = (\nona, \nofa)$ ($\DDb = (\nonb, \nofb)$) denotes
the corresponding on-off data.
This probability is determined by integrating the joint 
probability distribution of $\taua$ and $\taub$ over a relevant 
two-dimensional domain.
Here, a constant $A$ is used to account, at least to first order, 
for different observational conditions or experimental 
imperfections (see below).

From a practical perspective, the best way is to compare 
source fluxes. 
For this, we utilize the unconditional distributions of the differences 
$\deltaa$ and $\deltab$, respectively, see~\eqb{Q03}-\eqc{Q04}.
The probability that the flux $\ja = \deltaa / \eona$ observed 
in the first observation is less than the flux $\jb = \deltab / \eonb$ 
deduced from the second one, both fluxes treated as
random variables, is
\beql{Q13}
P( \ja < \jb ) = 
P ( \deltaa < \frac{\eona}{\eonb} \deltab ) = 
\Iint{-\infty}{\infty} \fdelaa(\xxaa)
\left[ \Iint{\frac{\eonb}{\eona} \xxaa}{\infty} 
\fdelbb(\xxbb) \dif{\xxbb} \right] \dif{\xxaa}.
\eeq
Here, the assessment of stability of source fluxes requires 
the knowledge of the on-source exposures, $\eona$ and $\eonb$. 
However, they may be affected by various imperfections 
associated with details of detection and data processing, 
especially when different sources are examined by different 
techniques.

The discrepancy between two independent observations can also 
be described by comparing the ratio variables while canceling 
out the effect of exposures.
If we adopt the unconditional distributions for the ratio 
variables $\betaa$ and $\betab$ given in~\eqa{Q06},
the inconsistency between two sets of on-off data can be quantified 
by the probability 
\beql{Q14}
P( \betaa < \xi \betab ) = 
\Iint{0}{\infty} \fbetaa(\xxaa)
\left[ \Iint{\xi^{-1} \xxaa}{\infty} 
\fbetbb(\xxbb) \dif{\xxbb} \right] \dif{\xxaa}.
\eeq
Here, for further possible applications, we introduced a parameter 
$\xi > 0$, allowing us to compare multiples of the ratio variables. 
In a first order approach, this parameter can be employed to eliminate
imperfections attributable to detection and data evaluation.

When two measurements collected in the same on- and off-source
zones are studied ($\alphaa = \alphab$), the proportion $\omega$ 
is advantageously used after a straightforward modification 
of~\eqa{Q14}.
Note also that the proposed probabilities are easily modified if sources 
are assumed to be present in their on-source zones, see~\app{App03}. 
Specifically, when non-negative source rates are guaranteed due 
to external arguments, the probabilities of inconsistency are obtained 
by putting the conditional distributions into the relevant equations 
while changing the integration limits accordingly. 

The integration in~\eqb{Q13} and~\eqc{Q14} is to be performed
numerically over the indicated two-dimensional sets. 
In some counting experiments, background rates can be estimated 
with sufficient accuracy from auxiliary measurements or modeled 
numerically.
With this simplification, we obtained explicit formulae for 
the probabilities of inconsistency summarized in~\app{App05}. 

The probabilities of inconsistency given in~\eqb{Q13}~and~\eqc{Q14}
have somewhat different meanings.
The difference $\delta$ allows us to quantify disparities between 
source fluxes, when on-source exposures are known. 
The probabilities based on the fractional variables $\beta$ and $\omega$ 
describe discrepancies between on-source observations when 
expressed with respect to the background or total measurements, 
respectively.
Thus, in more complicated cases, additional information about details 
of detection and data processing is needed for their correct interpretation 
(e.g. background rates, energy ranges, data quality limits etc.). 

The probabilities written in~\eqb{Q13}~and~\eqc{Q14} do not substitute 
for the probabilities of the source presence in the on-source zone, 
see~\app{App01}.
Indeed, it can be more likely that a larger flux is observed from 
a source which is found to be less significant than the other, 
i.e. $P(\ja < \jb) > 0.50$ while $P_{1}^{+} > P_{2}^{+}$ 
and vice versa.
Note also that quantified disparities between source fluxes, 
$P(\ja < \jb)$, when compared to ratio results, 
$P(\betaa < \betab)$, for a given pair of observations,  
may reveal hitherto unnoticed features that could affect measurements, 
were not considered during data processing or disrupted homogeneity 
of the underlying Poisson processes.

%------------------------------------------------------------------
\section{Examples}
\label{Sec03}

The usefulness of the method described in~\sct{Sec02} is demonstrated 
using arrival directions of the highest energy cosmic rays measured 
by the Pierre Auger Observatory~\cite{Aug03,Aug04,Aug05}.
Considering a predefined set of positions of nearby 
active galactic nuclei (AGN), we provide information to what extent 
is this set of possible sources related to directional data
after this association has been suggested~\cite{Aug01,Aug02}.
In a similar way, we also examine a signal that has been initially 
associated with the region around Centaurus~A (Cen~A)~\cite{Aug03,Aug04}.
We emphasize that earlier conclusions~\cite{Aug01,Aug02,Aug03,Aug04,Aug05} 
are in line with our analysis. 
Our aim is not to reassess previous studies, we only point out how 
the previous findings may be viewed from different perspectives.

Regardless of the results of further analysis~\cite{Aug03,Aug05}, 
we assumed that the signals from AGNs~\cite{Aug01,Aug02} and 
Cen~A~\cite{Aug03,Aug04} have not yet been confirmed. 
Given the data that were observed in the preselected on-off regions, 
we calculated the posterior distributions of the difference and 
fractional variables. 
We assumed the same prior distributions for the on- and off-source 
means with common shape parameters and zero rate parameters, 
i.e. $s = \ssp = \ssq$ and $\gamma = \gp = \gq \to 1$.
Furthermore, we derived the posterior distributions of the source 
flux $j$ using $j = \delta / a$ where $a = \aon A$ and $A$ denotes 
the integrated exposure of the period of data taking. 

In the following, we show how the three on-off variables
can be used when examining the previously suggested associations. 
Based on the results of~\sct{Sec02b}, we provide examples related 
to the issue of waiting for the next on-source events.
We also present examples of how to compare various independent 
measurements, see~\sct{Sec02c}. 
In the latter case, we include the latest hot spot (HS) data obtained 
by the Telescope Array surface detector~\cite{Tar03}. 
%------------------------------------------------------------------
\begin{table}[ht!]
\vspace{-0.5cm}
%\rotatebox{90}{\begin{varwidth}{\textheight}
\caption{\small 
AGN and Cen~A data measured by the Auger surface 
detector~\cite{Aug03,Aug04,Aug05}
and the HS data detected by the Telescope Array~\cite{Tar03}. 
Source assignment, period, exposure $A$ in $\kmSsry$, measured on- 
and off-source counts and the on-off parameter $\alpha$ are listed 
in the first sixth columns.
The endpoints of examined periods are denoted by
$A =$ (May~27, 2006), $B =$ (Aug~31, 2007), $C =$ (Dec~31, 2009), 
or $C =$ (Jan~1, 2010) for Cen~A, and $D =$ (Mar~31, 2014), 
respectively.
For the HS we used the two-year data collected from 
$E =$ (May~5, 2013) to $F =$ (May~11, 2015), 
see Table~1 in Ref.~\cite{Tar03}.
The Bayesian probabilities of no source ($\Pm$), corresponding 
significances ($\SBa$) and Li-Ma significances ($\SLM$) are given
in the next three columns.
For Bayesian results, Jeffreys' prior distributions were adopted,
i.e. $s=\frac{1}{2}$ and $\gamma \to 1$.
\vspace{-0.5cm}
}
\begin{center}
%{\tiny
%{\scriptsize
{\footnotesize
\begin{tabular}{ l l r r r r r r r}
\\\hline\hline\\[-2mm]
Data & Period & A & 
$\non$  & $\noff$ & $\alpha$ & $P^{-}$ & $\SBa$ & $\SLM$ \\[2mm]
\hline\hline\\[-2mm]
AGN & $\AvB$ & 4500 & 
9 &  4 & 0.266 & $8.2 \, 10^{-5}$ & 3.77 & 3.73 \\
AGN & $\AvC$ & 15980 & 
21 & 34 & 0.266 & $1.7 \, 10^{-3}$ & 2.93 & 2.90 \\
AGN & $\AvD$ & 47363 & 
41 & 105 & 0.266 & $2.0 \, 10^{-2}$ & 2.05 & 2.03 \\[2mm]
\hline\\[-2mm]
Cen~A & $\CvD$ & 31383 & 
3 & 76 & 0.047 & 0.59 & -0.22 & -0.31 \\[2mm]
\hline\\[-2mm]
HS & $\EvF$ & & 
5 & 32 & 0.075 & $7.0 \ 10^{-2}$ & 1.48 & 1.40 \\[2mm]
\hline\hline\\[-2mm]
\end{tabular}
}
\end{center}
\label{T01}
%\end{varwidth}}
\end{table}
%------------------------------------------------------------------
%------------------------------------------------------------------
%------------------------------------------------------------------
%------------------------------------------------------------------
\begin{figure}[ht!]
\wse
\includegraphics*[width=0.99\linewidth]{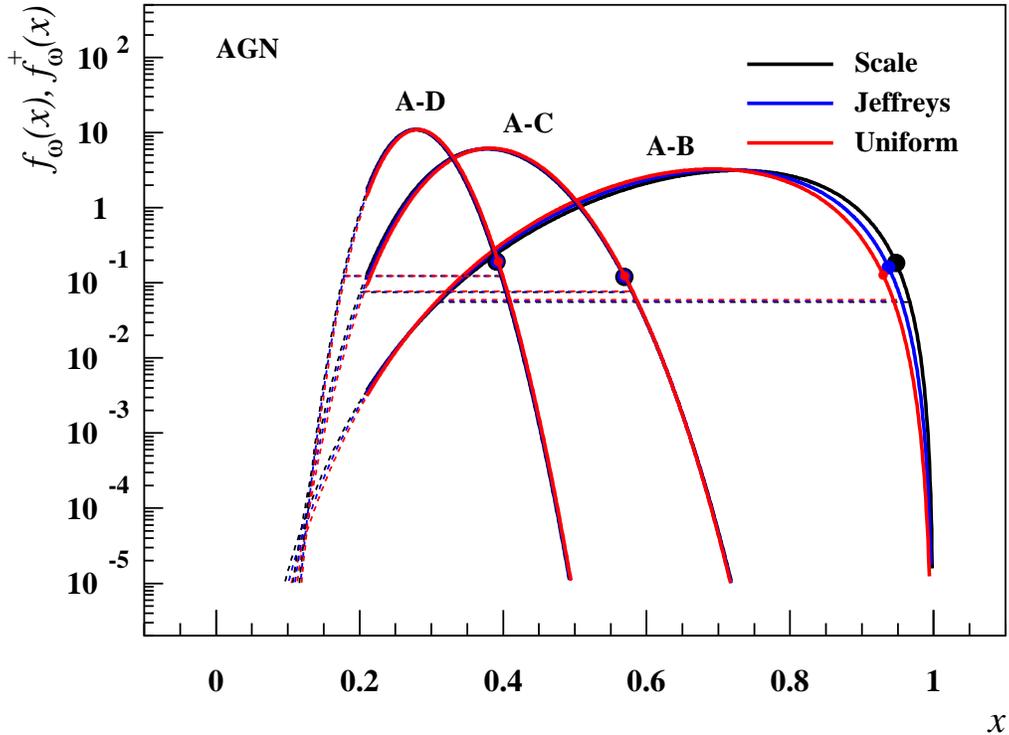}
\wsa
\caption{\small
Distributions of proportion $\omega$ for AGN data~\cite{Aug05}. 
The same uninformative priors for on- and off-source means 
($s= \ssp = \ssq$ and $\gp=\gq \to 1$) are used. 
Results for scale invariant ($s \to 0$), Jeffreys' ($s = \frac{1}{2}$) 
and uniform ($s = 1$) priors are shown in black, blue and 
red, respectively.
Distributions for the proportion, $\fom(x)$, and 
distributions $\fomp(x)$, when conditioned on a non-negative 
source rate ($\omega \ge \aon$), are depicted 
as dashed and thick full curves, respectively.
Horizontal dashed lines visualize credible intervals for the proportion 
($\langle \omml , \ommr \rangle$) at a $3 \sigma$ level of confidence. 
Upper limits at the same confidence level for the proportion assumed to 
be non-negative ($\ommr^{+}$ for $\omega \ge \aon$) are shown by colored 
points.
}
\label{F01}
\end{figure}
%------------------------------------------------------------------
%------------------------------------------------------------------
\begin{figure}[ht!]
\wse
\includegraphics*[width=0.99\linewidth]{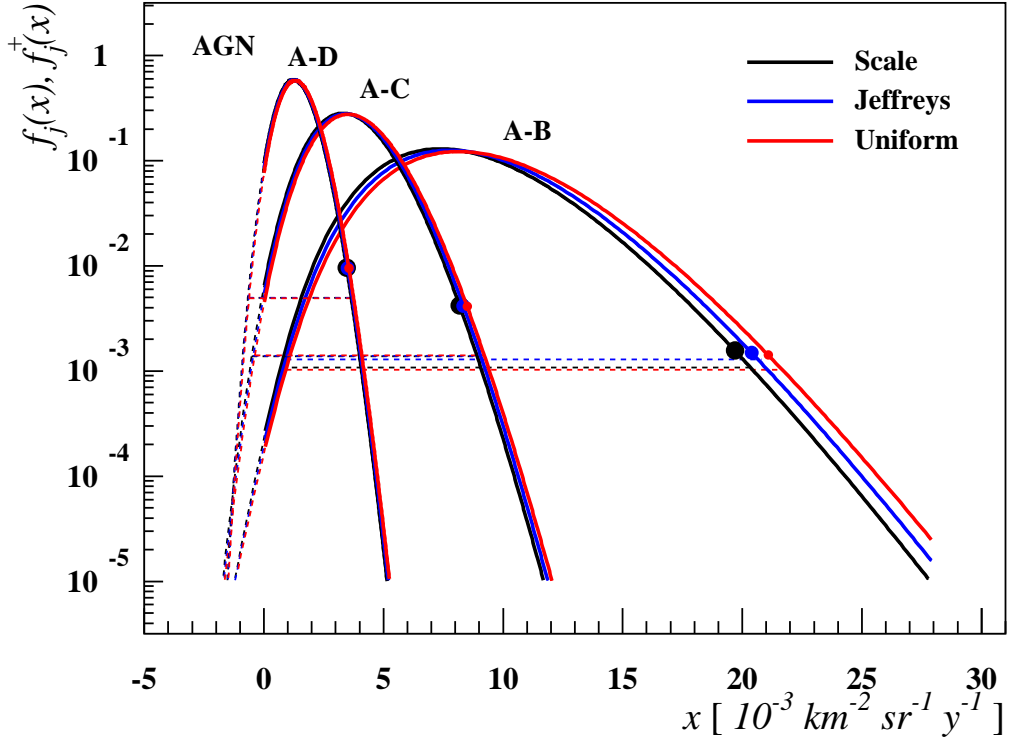}
\wsa
\caption{\small
Distributions of source flux $j = \delta / a$ ($a = \aon A$)
for AGN data~\cite{Aug05}. 
Both types of distributions are shown, $\fjj(x)$ (dashed curves) and 
$\fjjp(x)$ for $j \ge 0$ (thick full curves). 
For further details see caption to~\fig{F01}.
}
\label{F02}
\end{figure}
%------------------------------------------------------------------
%------------------------------------------------------------------
\begin{figure}[ht!]
\wse
\includegraphics*[width=0.99\linewidth]{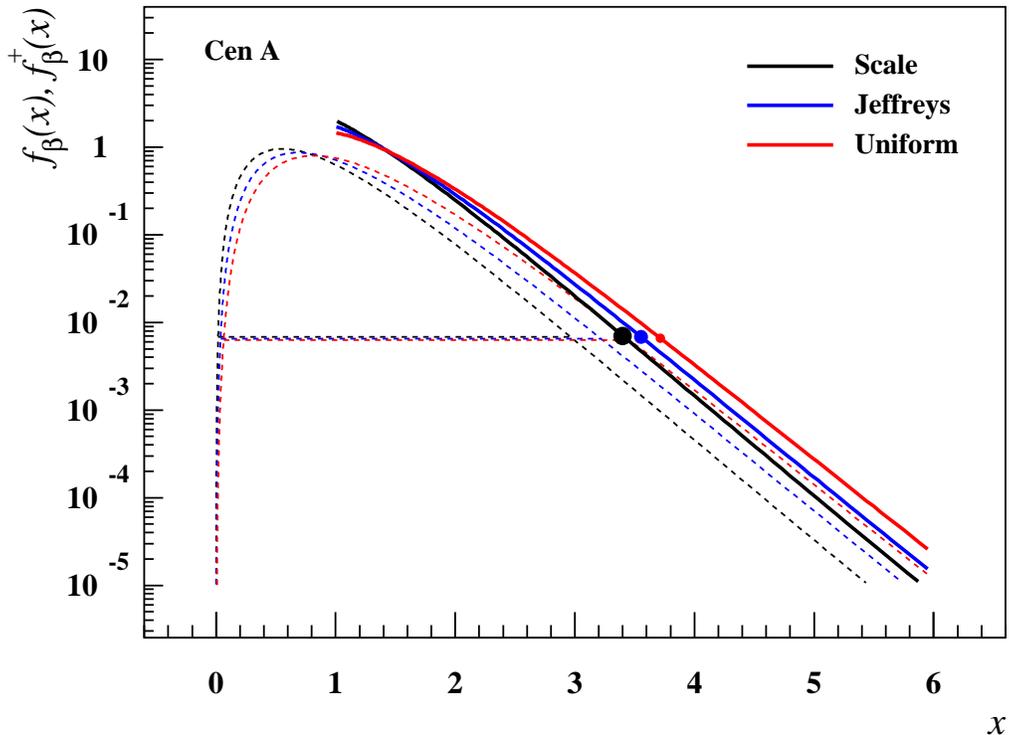}
\wsa
\caption{\small
Distributions of ratio $\beta$ for Cen~A data~\cite{Aug05}. 
Both types of distributions are shown, $\fbet(x)$ (dashed curves) 
and $\fbetp(x)$ for $\beta \ge 1$ (thick full curves).
For further details see caption to~\fig{F01}.
}
\label{F03}
\end{figure}
%------------------------------------------------------------------
%------------------------------------------------------------------
\begin{figure}[ht!]
\wse
\includegraphics*[width=0.99\linewidth]{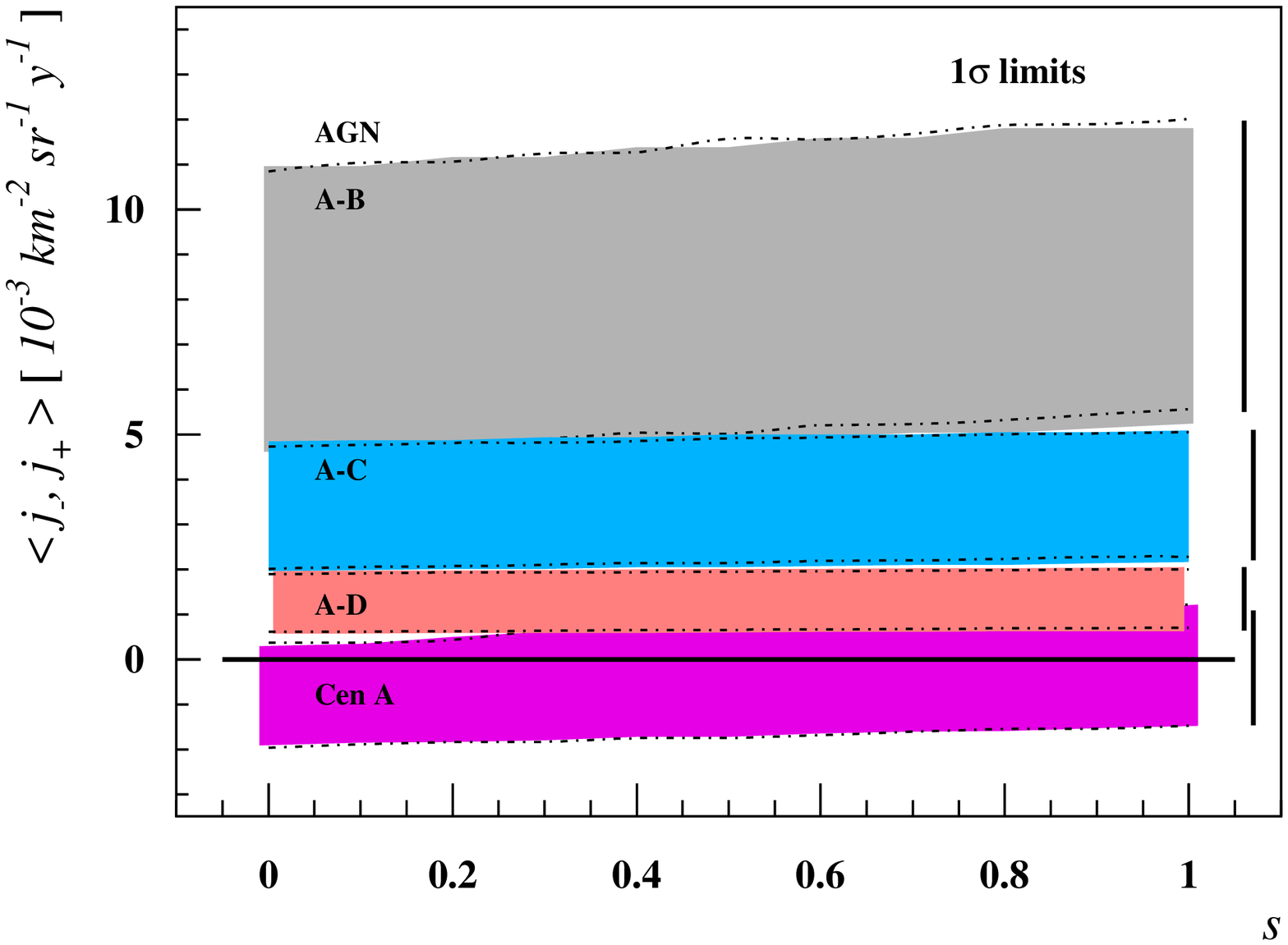}
\wsa
\caption{\small
Credible intervals for source flux $j = \delta / a$ ($a = \aon A$) 
at a $1 \sigma$ level of confidence are shown as functions of 
the common shape parameter of prior distributions $s = \ssp = \ssq$
($\gp = \gq \to 1$). 
Results for AGN (gray, blue and red bands) and Cen~A (magenta) 
data~\cite{Aug05} are depicted.
Dashed-dot lines indicate limits estimated using the approach based 
on known background (see~\app{App04}).
Black vertical lines show classical limits deduced within the unbounded 
profile likelihood analysis~\cite{Rol01}.
The horizontal black line represents the background expectation.
}
\label{F04}
\end{figure}
%------------------------------------------------------------------
%------------------------------------------------------------------
\begin{figure}[ht!]
\wse
\includegraphics*[width=0.99\linewidth]{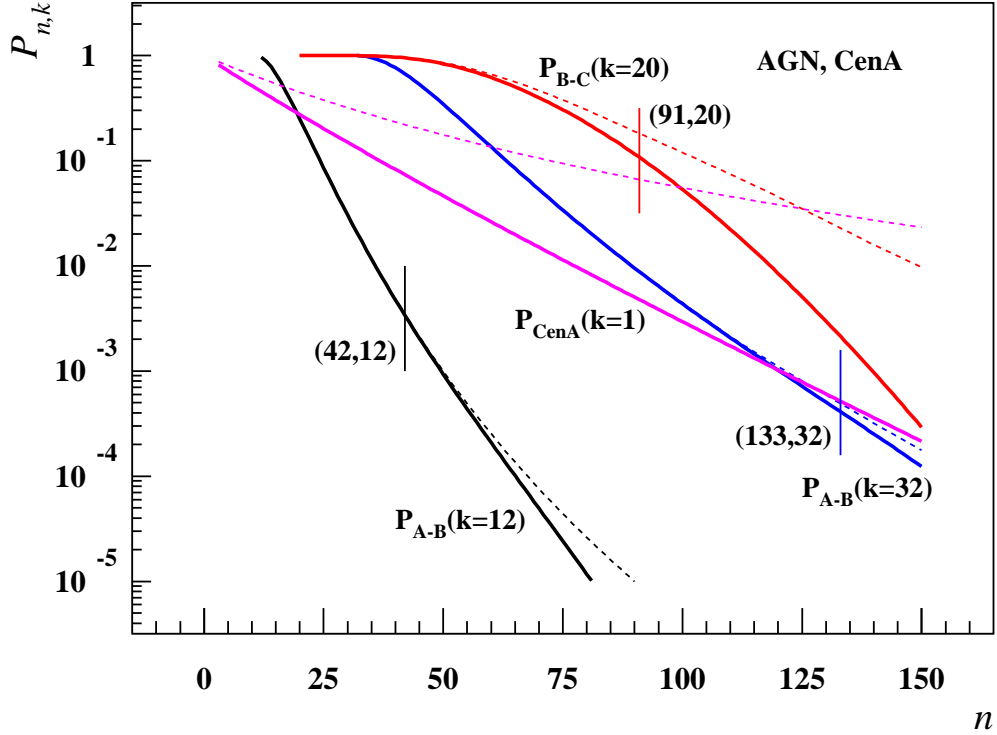}
\wsa
\caption{\small
Waiting time predictions.
The probabilities $P_{n,k}$ that less than $k$ on-source 
events are observed are shown as functions of the total number 
of registered events $n$.
Predictions based on the AGN signals observed in $\AvB$ period 
for the next 12 (32) AGN events are shown in black (blue). 
$\BvC$ predictions for the next 20 AGN counts are in red.
Magenta lines are for predictions of one next Cen~A event, 
based on the Cen~A data from $\CvD$ period.
Dashed (full) lines show unconditional (conditional) results 
based on Jeffreys' prior distributions.
Colored vertical lines indicate observations of $(n, k)$ events 
collected in the subsequent AGN periods. 
}
\label{F05}
\end{figure}
%------------------------------------------------------------------
%------------------------------------------------------------------
\begin{figure}[ht!]
\wse
\includegraphics*[width=0.99\linewidth]{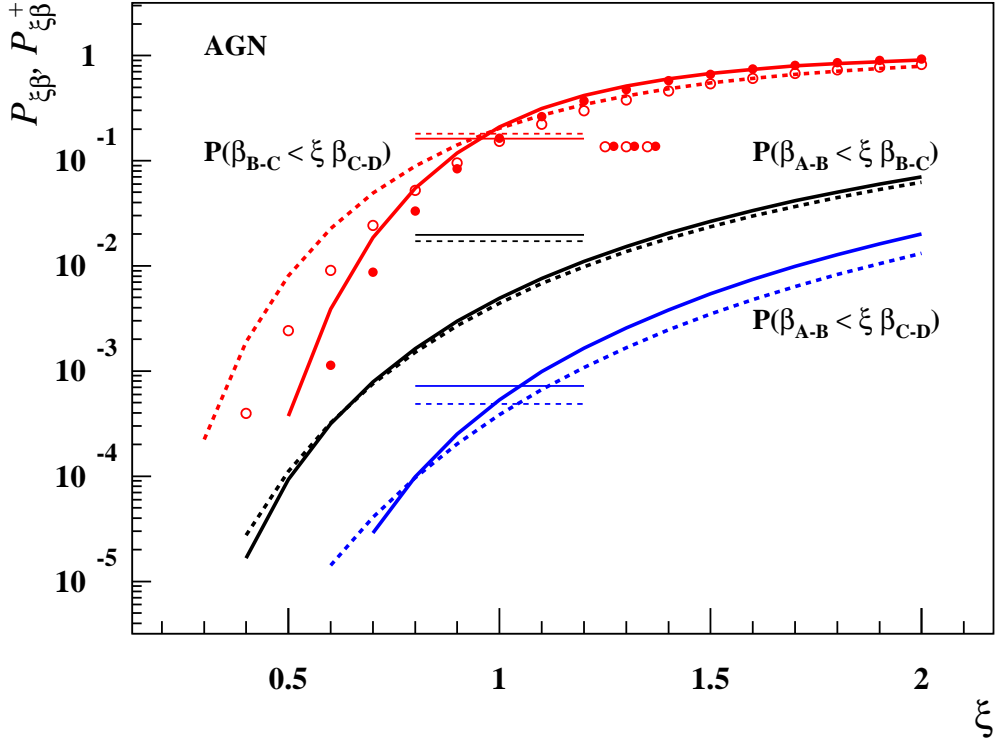}
\wsa
\caption{\small
Probabilities of inconsistency for the ratio $\beta$,
$P_{\xi\beta} = P(\betaa < \xi \betab)$,
deduced from the AGN data are shown as functions of the 
parameter $\xi$ (see~\sct{Sec02c}). 
Black, blue and red lines are for the comparison of three separated 
AGN periods.
Dashed and full lines show unconditional ($P_{\xi \beta}$) and conditional 
($P_{\xi \beta}^{+}$) results, respectively, based on Jeffreys' priors. 
Red empty (full) points show unconditional (conditional) results
for $\BvC$ and $\CvD$ periods assuming known background rates 
(see~\app{App04}) and uniform priors for on-source 
means.
Thin horizontal lines indicate the probabilities of inconsistency,
$P( \ja < \jb )$, between AGN fluxes.
Horizontal chains of three red points are for source fluxes provided 
that background rates are known (see~\app{App05}).
}
\label{F06}
\end{figure}
%------------------------------------------------------------------
%------------------------------------------------------------------
\begin{figure}[ht!]
\wse
\includegraphics*[width=0.99\linewidth]{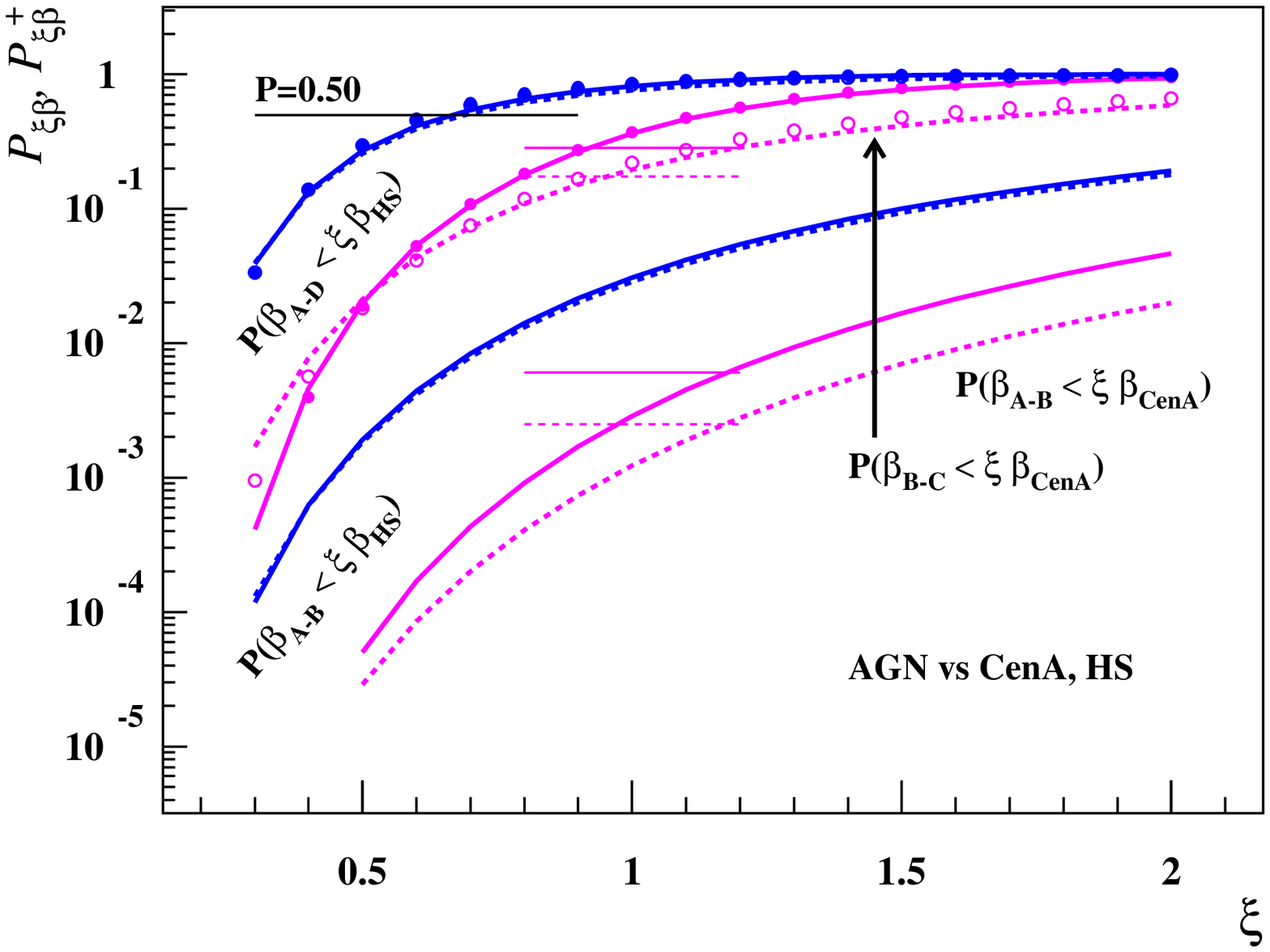}
\wsa
\caption{\small
Probabilities of inconsistency for the ratio $\beta$ are shown 
as functions of the parameter $\xi$ (see~\sct{Sec02c}). 
The AGN data collected in periods $\AvB$ and $\BvC$ are 
compared to the Cen~A signal in period $\CvD$, see magenta 
lines.
The probabilities that quantify inconsistency between the HS signal 
and the AGN data registered in periods $\AvB$ and $\AvC$ are shown 
in blue.
The black horizontal line indicates a probability of 0.50. 
For further details see caption to~\fig{F06}.
}
\label{F07}
\end{figure}
%------------------------------------------------------------------

%------------------------------------------------------------------
\subsection{Active galactic nuclei}
\label{Sec03a}

Among other important results~\cite{Aug05}, one of the topics 
of discussion regarding the distributions of arrival directions 
of the highest energy cosmic rays has focused on their association 
with a set of positions of nearby objects from the 12th edition 
of quasars and AGNs~\cite{VCV01}.
An initially revealed signal~\cite{Aug01,Aug02} has been reinvestigated 
in subsequent studies using the newly registered data~\cite{Aug03,Aug05}.

In order to document the uses and advantages of the Bayesian reasoning, 
we examined data registered by the surface 
detector of the Pierre Auger Observatory since May~27, 2007 up to 
March~31, 2014 (see Table~A1 in Ref.~\cite{Aug05}), 
after the AGN signal was recognized~\cite{Aug01,Aug02}. 
Specifically, we used events with energies in excess of $53 \EeV$ and
with zenith angles not exceeding $60\dg$.
For the association of the selected events with the nearby AGNs 
we accepted a set of parameters as defined in Refs.~\cite{Aug01,Aug02} 
and then slightly modified~\cite{Aug03,Aug05}.
A complex AGN source consists of a unification of circular zones 
with angular radii $3.1\dg$ around the positions of AGNs within 
$75 \AMpc$ (redshifts $z \le 0.018$)~\cite{VCV01}.

We examined three sets of data collected successively, as reported 
in Refs.~\cite{Aug03,Aug05}.
Namely, we analyzed arrival directions of events registered since
May~27, 2006 up to August~31, 2007 (here denoted as period $\AvB$, 
II in Ref.~\cite{Aug03}), up to December~31, 2009 (here $\AvC$, II+III 
in Ref.~\cite{Aug03}) and, finally, up to March~31, 2014 (here $\AvD$, 
see also Ref.~\cite{Aug05}).
The integrated exposures of the Auger surface detector, measured counts 
of on- and off-source events and on-off parameters $\alpha$, 
all taken from Refs.~\cite{Aug03,Aug05}, are summarized in the first 
six columns in the upper three lines in~\tab{T01}.
In this table, we also show some statistical characteristics 
based on the Jeffreys' priors ($s = \frac{1}{2}$, $\gamma \to 1$) 
and asymptotic Li-Ma significances~\cite{Lim01}. 

The posterior distributions for the proportion $\omega$ are 
depicted in~\fig{F01}. 
In this figure, we show results with three kinds of uninformative 
prior distributions, namely, for scale invariant ($s \to 0$, in black), 
Jeffreys' ($s=\frac{1}{2}$, blue) and uniform ($s=1$, red) prior 
distributions.
Two families of posterior distributions are depicted, unconditional
distributions (dashed curves) as well as distributions conditioned on 
a non-negative source rate in the on-source region (thick full lines),
i.e. assuming $\omega \ge \aon$, see~\app{App03}.
In~\fig{F01}, also credible intervals and upper limits for 
the proportion $\omega$ at a $3\sigma$ level of confidence are 
visualized (see~\app{App02}).  

As an alternative, in~\fig{F02} we show posterior distributions 
for the AGN flux $j = \delta / a$, 
given the on-off data in three examined period, and again using 
the three uninformative prior options. 
Relevant credible intervals at a $1\sigma$ level of significance 
are depicted in~\fig{F04} as functions of the common shape 
parameter.
The classical estimates~\cite{Rol01} and the results with 
known background rates (see~\app{App04}) are also shown 
in~\fig{F04}.

The posterior distributions shown in~\figs{F01} and~\figg{F02} 
clearly illustrate that the Bayesian inferences are only slightly 
dependent on the choice of uninformative prior distributions 
($s \in \langle 0, 1 \rangle$, $\gamma \to 1$) 
if the AGN source exhibits a sufficiently high activity, 
see also~\fig{F04}.
In such cases, due to large probabilities of the source presence 
in the AGN region, all conditional distributions approximately 
follow in their domains relevant unconditional distributions.
Furthermore, we learned how accessible information about 
the AGN source evolves with an increasing number of events 
recorded in the three successive sets of on-off data.
Our Bayesian estimates agree with the reported fractions 
of events associated with the AGN region and their downward 
trend~\cite{Aug03,Aug05}.

A decreasing AGN signal is also reflected in the predictions 
of the waiting time for the next on-source events when compared 
with future observations, see~\sct{Sec02b}. 
In~\fig{F05}, we show the probability that less than a given number 
of AGN events were detected in a number of subsequent measurements, 
while relying on previous observations.
For example, the Auger data collected in $\AvB$ period predicts 
that a total of $42$ events should be registered prior to the 
next $12$ AGNs events with a probability below $4 \ 10^{-3}$ 
(black lines).
Hence, when confronted with the Auger data from $\BvC$ period, 
in which these numbers were observed, such a waiting time is 
very unlikely. 
This result allows us to conclude that the $\BvC$ data is 
inconsistent with the $\AvB$ observation at about a $3\sigma$ 
level of confidence.

Independent AGN observations are compared in~\fig{F06}, see~\sct{Sec02c}.
In this case, the source fluxes as well as the ratios $\beta(\xi \approx 1)$ 
are well suited since still the same on- and off-source zones are observed 
with the same instrument. 
The parameter $\xi$ is employed to show the probability that one ratio 
is $\xi$-times smaller than the other or it can correct for
imperfections, if known (e.g. different background rates,
energy ranges, seasonal effects etc.). 
Our results depicted in~\fig{F06} agree with the findings drawn from 
the waiting time analysis. 
Namely, it is very unlikely that the AGN ratio from $\AvB$ period 
is less than the ratios derived from the two subsequent periods, 
and the same holds for the fluxes (black and blue results).
But the probability of inconsistency between $\BvC$ and $\CvD$ periods 
are much larger (red results).
Note also that the discrepancy between the probabilities calculated 
for the AGN fluxes and ratios, when relating $\AvB$ and $\BvC$ 
periods for $\xi \approx 1$ (in black), 
may indicate inhomogeneities of the underlying Poisson processes.

We also examined two possible signals deduced from different 
on-off measurements that were collected by different experiments.
In~\fig{F07}, the two-years HS data collected on the northern 
hemisphere by the Telescope Array surface detector (see Table~1 
in Ref.~\cite{Tar03}) is compared to the AGN signal measured 
by the Auger surface array on the southern 
hemisphere~\cite{Aug03,Aug05}.
Using two sets of the AGN data, $\AvB$ and $\AvD$ periods, 
the probabilities of inconsistency for the ratio $\beta$ are shown 
as functions of the parameter $\xi$ (blue lines).
Interestingly, since 
$P(\beta_{\rm A-D} < \xi \beta_{\rm HS}) > 0.70$ for $\xi \approx 1$, 
it is more likely that the less visible HS source 
($\SBa = 1.48$, see~\tab{T01}) manifests itself more markedly, 
when confronted with background, than the latest signal 
from AGN emitters ($\AvD$ period) which are more easily identified 
($\SBa = 2.05$, see~\tab{T01}).
In this example, the parameter $\xi$ may be utilized to correct for 
different energy scales ($E \ge 55 \EeV$ for the HS~\cite{Tar03} while 
$E > 53 \EeV$ for the AGNs~\cite{Aug05} plus systematic uncertainties) 
and for different background fluxes (at these energies, 
the overall flux on the northern hemisphere was measured to be at least 
twice as large as the southern flux, see e.g. Ref.~\cite{Aug06}).
If the northern background is truly larger than the southern one 
and, consequently, the observation of the HS signal is more difficult, 
one can correct for this effect by using $\xi > 1$, enlarging even more
the probability that the AGNs are weaker emitters. 

%------------------------------------------------------------------

%------------------------------------------------------------------
\subsection{Centaurus~A}
\label{Sec03b}

Centaurus~A (NGC 5128), located at a distance less than $4 \AMpc$, 
is known as a promising candidate source of the highest energy 
cosmic rays.   
Moreover, the nearby Centaurus cluster with large concentration 
of galaxies lies in approximately the same direction, at a distance 
of about $50 \AMpc$.
The excess of the highest energy events found in the vicinity of Cen~A 
and the properties of observed signal have been originally reported 
in Ref.~\cite{Aug03}.
However, this observation was not confirmed in successive 
measurements~\cite{Aug05}.

In this example, we show how the disappearance of a previously 
specified signal~\cite{Aug03} can be justified by using subsequently 
collected data within the Bayesian analysis.
We adopted the data registered by the surface detector of the Pierre Auger 
Observatory since January~1, 2010 up to March~31, 2014~\cite{Aug05} 
(here period $\CvD$), after the original Cen~A signal was 
identified~\cite{Aug03}.
The arrival directions of events with energies above $53 \EeV$ and zenith 
angles up to $60\dg$ were taken from Table~A1 in Ref.~\cite{Aug05}.
Based on the previous findings~\cite{Aug03}, we assumed a circular region 
with an angular radius of $18\dg$, located around the position of Cen~A 
($\alpha = 201.4\dg, \delta = -43.0\dg$).
The basic characteristics of the Cen~A region and the numbers of events 
collected in the examined period are summarized in the last but one row 
in~\tab{T01}.

In~\fig{F03}, we give an example of most unbiased information on 
the highest energy cosmic rays associated with the preselected 
Cen~A zone, which can be derived from the latest data~\cite{Aug05}.
In this figure, the posterior distributions for the ratio 
$\beta$ and corresponding credible intervals at a $3\sigma$ level 
of confidence are shown for three kinds of uninformative prior 
distributions.
We distinguish for unconditional distributions 
($\beta \ge 0$) and distributions conditioned on a non-negative 
source rate in the on-source region ($\beta \ge 1$.)
Credible intervals for the source fluxes $j$ at a $1\sigma$ level 
of confidence are depicted in~\fig{F04} as functions of the common 
shape parameter $s \in \langle 0,1 \rangle$ ($\gamma \to 1$).

The Bayesian inference indicates that the presence of the source 
in the originally selected Cen~A region is less likely than 
its absence therein when observations since 2010 are considered, 
i.e. $\Pm \ge 0.50$ ($\SBa \le 0$) for almost all prior options, 
for the Cen~A flux see~\fig{F04}. 
This conclusion agrees with the classical results based on 
asymptotic techniques, see~\tab{T01}.
Hence, the conditional distributions for the ratio $\beta$, 
$\fbetp(x)$ shown in~\fig{F03}, poorly reflect reality.

The absence of the signal registered in the Cen~A region 
in the latest observation can be quantified using the waiting 
time for one next Cen~A event, see~\sct{Sec02b}.
It is found in a marked difference between unconditional 
and conditional predictions that disqualifies the latter 
option, see~\fig{F05}.
Based on this data, over fifty events should be needed 
in order that the new one was identified in the Cen~A 
region at a $90\%$ level of confidence. 
Using the method of~\sct{Sec02c}, the same is documented in~\fig{F07}.
Namely, it is more likely that the four-years Cen~A signal is weaker
than the AGN activity measured in two preceding periods 
(magenta results). 
Here, the parameter $\xi$ can account for different background 
zones of Cen~A and AGNs emitters and different shapes of 
their energy spectra, for example.

In this regard, it is worth recalling that the Auger collaboration 
has lately pointed out that the significance of the excess of events 
in the angular windows and energy range, as examined in this study, 
is less than its originally observed value~\cite{Aug05}.
This was obtained by using a broader set of data collected between 
January~1, 2004 and March~31, 2014, including events with zenith angles 
up to $80\dg$, when the hypothesis of isotropy was tested.
The most significant departure from isotropy in the available set 
of data was reported for events with energies beyond $58 \EeV$ and 
with arrival directions within a circle of an angular radius of $15\dg$ 
centered on Cen~A~\cite{Aug05}.

%------------------------------------------------------------------

%------------------------------------------------------------------
\section{Conclusions}
\label{Sec04}

We focused on the search for new phenomena, when all relevant 
characteristics of a source which is suspected of causing observed 
effects cannot be set in an optimal way.
The issue was dealt with in the context of on-off measurements assuming 
registration of small numbers of events that obey Poisson distributions.
For this purpose, the Bayesian way of reasoning was utilized. 
This approach is not only statistically well justified and 
intuitively easily interpretable, but also provides readily 
computable results. 

We examined three appropriately chosen on-off variables that 
store information available from the on-off experiment.  
In addition to traditionally presented results, we proposed 
how to utilize observation-based information for predictions 
and comparisons, focusing on quantification of signal stability.

By using successive measurements, increasing sets of the highest 
energy events collected at the Pierre Auger Observatory were examined.
For comparison, also directional data reported by the Telescope Array 
was considered.
Using the recent Auger observations, we summarized the outputs
accessible in the proposed approach.
We discussed the extent to which the comparison of on-off 
measurements may help when searching for cosmic ray sources. 

%------------------------------------------------------------------
%\newpage
\vspace*{0.5cm}
{\bf Acknowledgments:}
We would like to acknowledge and thank our colleagues from the Pierre Auger 
Collaboration for many helpful discussions and, especially, for the tremendous 
work with data, its analysis and interpretation from which we benefit.
We thank anonymous reviewer for valuable comments and suggestions that helped 
us improve the presentation of this paper. 
This work was supported by the Czech Science Foundation grant 14-17501S.
The research of J.N. was partly supported by the Czech Science Foundation 
under project GACR~P103/12/G084.
\vspace*{0.5cm}

%
%\newpage
%------------------------------------------------------------------
\appendix
%\begin{appendices}
%\renewcommand\thesection{Appendix \Alph{section}}

%------------------------------------------------------------------
\section{Source detection}
\label{App01}

Using the posterior distribution for the difference, see~\eqb{Q03} 
and~\eqc{Q04}, the probability for the absence of a source 
in the on-source region is~\cite{Nos01}
\beql{A01}
\Pm = I_{\rrhb}(p,q), 
\eeq
where $p = \non+\ssp$, $q=\noff+\ssq$, $\rho = \gpqc$ 
and $I_{x}(a,b)$ denotes the regularized incomplete Beta 
function~\cite{Olv01}.
Using other on-off variables, we obtain after straightforward 
calculations
\beql{A02}
\Pm = P(\tau \le \atau) = \Iint{-\infty}{0} \fdel(x) \dif{x} = 
\Iint{0}{1} \fbet(x) \dif{x} = \Iint{0}{\aon} \fom(x) \dif{x},
\eeq 
where $\tau = \delta, \beta, \omega$, 
while $\atau = 0, 1, \aon$ denotes the balance value for the difference, 
ratio and proportion, respectively.
The probability of the presence of a source in the on-source zone
is $\Pp = 1 - \Pm$.
When viewed in terms of a normal variate with zero mean and 
unit variance, the probability $\Pm$ is converted to a Bayesian 
significance $\SBa$.

%------------------------------------------------------------------
\section{Credible intervals}
\label{App02}

For the shortest credible interval, $\langle \tau_{-}, \tau_{+} \rangle$, 
that contains the on-off variable with a probability $P$, 
one has to solve numerically ($\tau = \delta, \beta, \omega$)
\beql{B01}
P = \Iint{\tau_{-}}{\tau_{+}} \ftau(x) \dif{x}, 
\qqc
\ftau(\tau_{-}) = \ftau(\tau_{+}),
\eeq
under the indicated condition on endpoints of the corresponding 
probability density function $\ftau(x)$.
An upper bound for the source intensity, $\tau_{+}$, is determined 
numerically using the integral in~\eqa{B01} where we put 
$\tau_{-} \to -\infty, 0$ or $0$ for $\tau = \delta, \beta$ or $\omega$, 
respectively, and relax the indicated limitation on endpoints 
of $\ftau(x)$. 
For a non-negative source intensity ($\mon \ge \mb$), see~\app{App03}, 
credible intervals are derived by putting 
$\ftau(x) \to \ftaup(x)$ into~\eqa{B01} while we set
$0,1$ or $\aon \le \tau_{-} < \tau_{+} < +\infty,+\infty$ or $1$, 
respectively. 
For upper bounds for a know source we set directly $\tau_{-} = 0, 1$ 
or $\aon$ and relax the limitation on endpoints of $\ftaup(x)$. 

%------------------------------------------------------------------
\section{Known source}
\label{App03}

In a variety of problems we know with certainty that an active source 
is present in the on-source region or at least we have a good indication 
that it may be assumed. 
This issue is encountered when searching for accompanying radiation 
from already identified emitters, for example. 
When the mean event rate in the on-source zone can only increase 
beyond what is expected from background, the corresponding probability 
distributions are derived conditioning on the non-negative values 
of the difference of the on-source and background means, i.e. 
$\mon \ge \mb = \alpha \moff$ or, alternatively, $\tau \ge \atau$ 
where $\tau = \delta, \beta, \omega$ and $\atau = 0, 1, \aon$ 
for the difference, ratio and proportion, respectively.
For the conditional distributions we have
\beql{C01}
\ftaup(x) = \ftau(x \midd x \ge \atau) = \frac{ \ftau(x)}{\Pp}, \qqc
x \ge \atau,
\eeq 
where the Bayesian probability of the presence of a source in the 
on-source zone, $\Pp = 1 - \Pm = I_{\rrha}(q,p)$, follows from~\eqa{A01}.
Note that if the probability $\Pp$ approaches one, when it is 
exceedingly likely that the source contributes to the intensity 
detected in the on-source zone, the on-off problem is well described 
in the unconditional regime, since $\ftaup(x)$ tends to $\ftau(x)$ 
in the domain where $x \ge \atau$.

We recall that, by this construction, we obtain results which 
were derived in another way~\cite{Kno01,Cas01,Pro01,Pro02,Gil01,Gre01}, 
assuming that the source and background rates are non-negative, 
i.e. $\ms = \mon - \mb \ge 0$ and $\mb = \alpha \moff \ge 0$, 
for more information see Ref.~\cite{Nos01}.
Specifically, in the context of the on-off problem, the use of 
the proportion $\omega$ with the Jeffreys' prior distributions was 
advocated in Ref.~\cite{Gil01}.
In our scheme, substituting the corresponding parameters  
($p = \non + \frac{1}{2}$, $q = \noff + \frac{1}{2}$ and  
$\gp = \gq \to 1$) 
into~\eqb{Q08} and~\eqc{C01}, the posterior $\omega$-distribution 
written in Eq.(27) in Ref.~\cite{Gil01} is recovered.

%------------------------------------------------------------------
\section{Known background}
\label{App04}

The probability distributions of examined variables are further 
simplified in the case of a known background.
Such a simplification may be used, for example, when searching 
for sources of cosmic rays in a small on-source region  
($0 < \alpha \ll 1$) complemented by a much larger off-source zone 
which is comprised of the remaining part of the sky within 
the field of view of the experiment, where $\noff \gg 1$.
Then, the number of background events observed 
in the on-source zone follow approximately the Poisson 
distribution with an estimated mean parameter 
$\mb = \alpha \moff \approx \alpha \noff$, 
since its estimated variance is negligible, 
$\sigma^{2}(\alpha \noff) \approx \alpha^{2} \noff \ll \mb^{2}$. 
Another example is the analysis of a counting experiment that 
utilizes a constant background rate estimated based on modeling 
considerations.

In such a case, we easily obtain 
$\mon = (\delta + \mb) \sim \Gaa(p,\gp)$~\cite{Nos01} 
and the ratio $\beta = (\mon / \mb) \sim \Gaa(p,\gp \mb)$, 
where $p = \non+\ssp$.
The proportion is given by the transformation
$\omega = (\alpha \beta) / (1 + \alpha \beta)$.
In summary, the probability density functions of all on-off variables 
are, respectively,
\beql{D01}
\fdelb(x) = \frac{\gp^{p}}{\Gamma(p)} (x + \mb)^{p-1} e^{-\gp (x+\mb)},
\qqa
x \ge -\mb,
\eeq
\beql{D02}
\fbetb(x) = \frac{(\gp \mb)^{p}}{\Gamma(p)} x^{p-1} e^{-\gp \mb x},
\qqc
x \ge 0,
\eeq
and
\beql{D03}
\fomb(x) = \frac{(\gp \moff)^{p}}{\Gamma(p)} \frac{x^{p-1}}{(1-x)^{p+1}} 
e^{-\frac{\gp \moff x}{1-x}},
\qqa
x \in \langle 0, 1 \rangle.
\eeq

In addition, assuming non-negative source rate in the on-source region, 
$\mon \ge \mb$ 
(i.e. $\delta \ge 0$, $\beta \ge 1$ or $\aon \le \omega \le 1$), 
we have for the corresponding probability density functions
\beql{D04}
\ftaubp(x) = \frac{\ftaub(x)}{\Ppb}, \qqb
x \ge \atau,
\eeq
where $\tau = \delta, \beta, \omega$, while $\atau = 0, 1, \aon$, 
and $R^{+}$ is the probability of the presence of a source 
in the on-source region provided a constant background mean 
is used\footnote{
Note that there are typographical errors in Eqs.(26) and (27) 
in Ref.~\cite{Nos01}.
There should be $\Gamma(p,\gp \mb)$ instead of $\Gamma(p,\mb)$.}, 
i.e.
\beql{D05}
\Ppb = \Iint{0}{\infty} \fdelb(x) \dif{x} = 
\Iint{1}{\infty} \fbetb(x) \dif{x} = 
\Iint{\aon}{1} \fomb(x) \dif{x} = 
\frac{\Gamma(p,\gp \mb)}{\Gamma(p)},
\eeq 
where
$\Gamma(a,x) = \Iint{x}{\infty} t^{a-1} e^{-t} \dif{t}$ 
is the upper incomplete Gamma function. %\footnote{}
It is useful to know that 
$R^{+}(p,x) = \frac{\Gamma(p,x)}{\Gamma(p)} = 
e^{-x} \Ssum{k=0}{p-1} \frac{x^{k}}{k!}$ 
for integer values of $p$.

Notice that for $\gp \to 1$, $\Pmb = 1 - \Ppb$ is the $p$-value 
obtained in the classical framework, when the background hypothesis 
(i.e. $\mon \le \mb$) is tested against the alternative of a source 
presence in the on-source zone ($\mon > \mb$) for the Poisson 
sampling process~\cite{Cou02}.

%------------------------------------------------------------------
\section{Comparison with known backgrounds}
\label{App05}

When fluctuations in the background are completely disregarded, 
see~\app{App04}, the probabilities of inconsistency introduced 
in~\sct{Sec02c} can be expressed explicitly. 
We assume two independent observations, marked by indices 
$1$ and $2$.
If only non negative integer values of relevant shape parameters 
($\sspa$ and $\sspb$) are considered, the integration in~\eqa{Q14} 
is easily performed using the posterior distributions given 
in~\eqa{D01}.
Then, the probability of inconsistency between source fluxes 
when the on-source exposures ($\eona$ and $\eonb$) are known, 
see~\eqa{Q13}, can be written in a compact formula 
($p_{1} = \nona + \sspa, p_{2} = \nonb + \sspb$, $p_{1}, p_{2} \in N$)
\beql{E01}
P(\ja < \jb) = 
\frac{e^{-u} \ v^{p_{2}}}{(1+v)^{p_{1}+p_{2}}} 
\Ssum{k=1}{p_{2}} 
\Ssum{i=k}{p_{2}} 
{p_{1} + p_{2} - i - 1 \choose p_{2} - i}
\frac{ u^{i-k}}{(i - k)!} \left( \frac{1+v}{v} \right)^{i} R_{i}(v). 
\eeq
Here, $u =  \gpbb \mbb - v \ \gpaa \mba \ge 0$ depends on the known 
background rates, $\mba$ and $\mbb$, 
$v = ( \gpbb \eonb) / (\gpaa \eona)$ depends on the ratio of two on-source 
exposures, $\gpaa$ and $\gpbb$ denote the prior rates of the on-source 
means and $R_{i}(v) =1$ for the unconditional $\delta$-distributions 
given in~\eqa{D01}, while 
\beql{E02}
R_{i}(v) = 
\frac{R^{+}(p_{1} + p_{2} - i, (1+v) \gpaa \mba)}
{R^{+}(p_{1},\gpaa \mba) \ R^{+}(p_{2},\gpbb \mbb)}, 
\eeq
for the conditional $\delta$-distributions, see~\eqb{D04} 
and~\eqc{D05}.
In~\eqa{E01} we compare source fluxes provided $u \ge 0$. 
If $u < 0$, we simply exchange measurements, using 
$P(\ja < \jb) = 1 - P(\jb < \ja)$.

In a similar way and under the same conditions, we can compare two 
independent on-off measurements through the ratios $\betaa$ and 
$\betab$ when background uncertainties are not considered.
Using the parameter $\xi$, the probability of inconsistency 
between two ratios (see~\eqa{Q14}) is written ($p_{1}, p_{2} \in N$)
\beql{E03}
P(\betaa < \xi \betab) = 
\frac{w^{p_{2}}}{(1+w)^{p_{1}+p_{2}}} 
\Ssum{k=1}{p_{2}} 
{p_{1} + p_{2} - k - 1 \choose p_{2} - k} 
\left( \frac{1+w}{w} \right)^{k} R_{k}(w),
\eeq
where $w = \xi^{-1} (\gpbb \mbb) / (\gpaa \mba)$ and $R_{k}(w) = 1$ 
for the unconditional $\beta$-distributions (\eqa{D02}) and for 
the conditional ones (\eqb{D04} and~\eqc{D05}) it is written 
in~\eqa{E02}.
The formula in~\eqa{E03} holds for $\xi \le 1$. 
If $\xi > 1$, we use
$P(\betaa < \xi \betab) = 1 - P(\betab < \xi^{-1} \betaa)$.

%\end{appendices}
%------------------------------------------------------------------

%\end{linenumbers}



\vspace*{0.5cm}
%\newpage
%------------------------------------------------------------------

%------------------------------------------------------------------

\end{document}